\documentclass[acmsmall]{acmart}

\usepackage{amsmath,amsfonts}
 
\usepackage{amssymb}

\usepackage{wrapfig} % 浮动表格
\usepackage{bbding} %对勾叉号

\usepackage{lipsum}
\usepackage{tikz}
\usetikzlibrary{arrows.meta}
\usetikzlibrary{automata,positioning}

\usepackage{threeparttable}

\definecolor{tabred}{RGB}{230,36,0}%
\definecolor{tabgreen}{RGB}{0,116,21}%
\definecolor{taborange}{RGB}{250,124,30}%
\definecolor{tabbrown}{RGB}{171,70,0}%
\definecolor{tabyellow}{RGB}{251,253,169}%
\newcommand*{\vcorr}{%
  \vadjust{\vspace{-\dp\csname @arstrutbox\endcsname}}%
  \global\let\vcorr\relax
}% 

\usepackage{mathtools}

\usepackage{lscape}
\usepackage[figuresright]{rotating}
\usepackage[graphicx]{realboxes}
\usepackage{adjustbox}
\usepackage{caption}

\usepackage{subfigure}
\usepackage{caption}

\usepackage{colortbl} 
\usepackage{xfrac}

\usepackage{appendix}
\usepackage{float}

\usepackage{fbox} %use box
\usepackage{fancybox}%use other types of boxes
\usepackage{framed}

\usepackage{multirow}

\def\BibTeX{{\rm B\kern-.05em{\sc i\kern-.025em b}\kern-.08em
    T\kern-.1667em\lower.7ex\hbox{E}\kern-.125emX}}
    
\usepackage{CJKutf8}  

\usepackage{hyperref}

\usepackage{tabularx,makecell, caption}
\usepackage{enumitem} % to control Itemization spacing

\newcolumntype{L}{>{\arraybackslash}X}

\usepackage{multicol}
\usepackage{listings}
\usepackage{blindtext}
\usepackage{etoolbox,xstring,mfirstuc,textcase}
\usepackage{listings}

\usepackage{subfiles}
\usepackage[most]{tcolorbox}

\usepackage{xcolor}
\theoremstyle{plain}                
\theoremstyle{definition}       

\usepackage{siunitx}
\usepackage{comment}
\usepackage{indentfirst} 
\usepackage{framed} 

\usepackage[font=small,labelfont=bf,tableposition=top]{caption}
\usepackage{booktabs}
\usepackage{dashbox}

\usepackage{listings}
\usepackage{url}

\usepackage{color}
\usepackage{algorithm,algpseudocode} %% select one of them

\renewcommand\footnotetextcopyrightpermission[1]{} % removes footnote with conference information in first column
\begin{document}

\title{Dataset Obfuscation: Its Applications to and Impacts on Edge Machine Learning}

%=================================================
%author
%=================================================

%\begin{comment}  % priority

\author{Guangsheng Yu}
\orcid{0000-0002-6111-1607}
\affiliation{%
  \institution{Data61, CSIRO}
  \city{Sydney}
  \state{NSW}
  \country{Australia}
  \postcode{2121}
}
\email{saber.yu@data61.csiro.au}

\author{Xu Wang}
\affiliation{%
  \institution{FEIT, UTS}
  \city{Sydney}
  \country{Australia}}
\email{xu.wang-1@uts.edu.au}

\author{Caijun Sun}
\authornote{Caijun Sun and Ping Yu act as the corresponding authors.}
\affiliation{%
  \institution{Zhejiang Lab}
  \city{Hangzhou}
  \country{China}
}
\email{sun.cj@zhejianglab.com}

\author{Ping Yu*}
\affiliation{%
 \institution{Faculty of Computing, Harbin Institute of Technology}
 \city{Harbin}
 \country{China}}
\email{yuping0428@hit.edu.cn}

\author{Wei Ni}
\affiliation{%
  \institution{Data61, CSIRO}
  \city{Sydney}
  \state{NSW}
  \country{Australia}
}
\email{wei.ni@data61.csiro.au}

\author{Ren Ping Liu}
\affiliation{%
  \institution{FEIT, UTS}
  \city{Sydney}
  \country{Australia}}
\email{renping.liu@uts.edu.au}

%\end{comment}

%=================================================
%abstract
%=================================================

\begin{abstract}
Obfuscating a dataset by adding random noises to protect the privacy of sensitive samples in the training dataset is crucial to prevent data leakage to untrusted parties for edge applications.
We conduct comprehensive experiments to investigate how the dataset obfuscation can affect the resultant model weights - in terms of the model accuracy, Frobenius-norm (F-norm)-based model distance, and level of data privacy - and discuss the potential applications with the proposed Privacy, Utility, and Distinguishability (PUD)-triangle diagram to visualize the requirement preferences. Our experiments are based on the popular MNIST and CIFAR-10 datasets under both independent and identically distributed (IID) and non-IID settings. Significant results include a trade-off between the model accuracy and privacy level and a trade-off between the model difference and privacy level. The results indicate broad application prospects for training outsourcing in edge computing and guarding against attacks in Federated Learning among edge devices.
\end{abstract}




% %=================================================
\maketitle
% %=================================================

\section{Introduction}
\label{sec_intro}

% \subsection{Background}
Artificial Intelligence (AI) has evolved for decades and the capability of machine learning has become one of the core components in many applications in human daily life, including edge computing networks~\cite{edge-1, edge-2}. Recently, formulating the AI ethics and building Responsible AI have been attracting increasing attention from industry and academia to ensure processes being legitimate, secure, and ethical for everyone involved. 

Data privacy is one of the hotspots with various privacy-preserving solutions being proposed in terms of designing privacy-aware machine learning algorithms and platforms~\cite{DLDP2016CCS, PPDP2015ACM, hesamifard2018privacy, 197247} or implementing cryptographic algorithms, such as homomorphic encryption, to encrypt the dataset~\cite{cryptoeprint:2014/331} prior to data disclosure. 
However, the privacy-aware machine learning algorithms address only the ``hidden trees in a forest'' problem by obfuscating the query results, and cannot prevent data leakage for each piece of data samples. On the other hand, homomorphic encryption is computationally intensive ~\cite{10.1145/3214303} and prohibitive for edge applications, and has to be carefully crafted with respect to different learning algorithms with weak generality~\cite{noisy-dataset}.

Obfuscating a dataset using random noises recently came to the scene due to the increasing importance of data privacy in edge networks~\cite{edge-privacy}, particularly the training dataset privacy, in machine learning processes where edge-edge or edge-cloud dataset sharing is essential. A general practice of preserving data privacy is to add random noise to datasets during the data sharing to preserve the data privacy as one needs no compulsory presumptions about the trust between him/her and the data receivers. 
The most relevant study evaluates only the impacts of obfuscation on model prediction accuracy~\cite{noisy-dataset}, with no mention of either the model distance or the data utility which can outweigh the accuracy in many application scenarios. The F-norm-based model distance has been widely used, and its importance has appeared in many areas.
A typical example is applying Proof-of-Learning (PoL)~\cite{9519402} in Federated Learning (FL), where the Frobenius-norm (F-norm) is used to evaluate the model distance between two versions of the model originating from the raw and obfuscated datasets, respectively. Another example is the Multi-Krum aggregation~\cite{multi-krum} in FL can be byzantine-fault-tolerant by removing the model outliers in terms of F-norm. Therefore, we consider assessing the F-norm-based model distance under dataset obfuscation, which has yet been learned in existing studies, is essential and can help to understand the nature of dataset obfuscation in the above applications, thus encouraging the advancement of privacy-preserving protocols.

The goal of this paper is to comprehensively investigate the impacts of noise-based dataset obfuscation on the trained models.
In this paper, we extend comprehensive experiments regarding the dataset obfuscation in machine learning, to analyze in-depth the impacts on the resultant models and explore potential application scenarios in edge networks.
The key contributions of this paper are as follows.

\begin{enumerate}
\item We determine a generic obfuscation model with various metrics considered over a training dataset.

\item We carry out various types of experiments over the MNIST~\cite{726791} and CIFAR-10~\cite{krizhevsky2009learning} datasets under both independent and identically distributed (IID) and non-IID conditions with a wide range of noise settings.

\item We discuss the challenges and potential applications in edge networks based upon the experimental results, including training outsourcing and Federated Learning (FL), with a new visualization tool, named Privacy, Utility, and Distinguishability (PUD)-triangle diagram.
\end{enumerate}

The experiments show that the model accuracy declines with the increase of noise level.
The increasing noise level also enlarges the model difference between feeding a raw dataset and its obfuscated version, thus weakening the training reproducibility.
Also, the disguisability of fake datasets becomes stronger with the increase of noise level due to the subequal model differences. The resultant observation indicates the existence of a balanced stable point at which all metrics can be acceptable for real applications, including training outsourcing on edge computing and guarding against attacks in FL among edge devices.

The rest of the paper is organized as follows.
The related works are given in Section~\ref{sec_relatedwork}.
Section~\ref{sec_system} provides the background.
The experiment settings and results are shown in Section~\ref{sec_experiments}, followed by discussions in Section~\ref{sec_discussions}. Section~\ref{sec-conclu} concludes the paper.
\section{Related Work}
\label{sec_relatedwork}

% Deep learning based on artificial neural networks has attracted wide attention in modeling, classifying and recognizing complex data. The amount of data is directly related to the success of model training. In this case, the exposure of data is a concern of deep learning participants. In \cite{PPDP2015ACM}, the authors designed a system that enables multiple parties to jointly learn an accurate neural network model without sharing their input datasets. The participants train independently on their own datasets and selectively share small subsets of their parameters during training process. The authors measured the relationship between model accuracy and the proportion of shared parameters. 

% The paper \cite{DLDP2016CCS} presented a machine learning framework TensorFlow for training models with differential privacy. The authors evaluated the approach on a popular image classification task by using the MNIST dataset.
% In \cite{hesamifard2018privacy}, the authors provided a framework CryptoDL to provide solutions for applying deep neural network algorithms to encrypted data. The approach train models using encrypted data, make encrypted predictions, and return the predictions in an encrypted form. The authors adopt polynomial approximation method to work within homomorphic encryption constrains.

Protecting privacy in machine learning has attracted wide attention and spawned many approaches in terms of collaborative learning where datasets are locally used at edge devices with only small subsets of parameters shared~\cite{PPDP2015ACM}, obfuscating the local model parameters by using random noise or differential privacy at edge devices~\cite{8843942}, adopting homomorphic encryption to encrypt data samples~\cite{hesamifard2018privacy}, etc.

This paper focuses on protecting dataset privacy in tasks where sharing datasets is essential between edges or between edge and cloud.
Authors of \cite{noisy-dataset} focus on the training outsourcing and propose an OBFUSCATE function to obfuscate training datasets before the disclosure. The function adds random noises to data samples or augments datasets with new samples to protect the dataset privacy. The authors also show that training upon the obfuscated dataset can preserve the model accuracy. 
Our paper conducts more comprehensive experiments to investigate impacts on not only the model accuracy, but also the model difference under both IID and non-IID settings (the latter of which is yet to be considered in existing studies). We explore and discuss real-world applications in edge computing which can take advantage of dataset obfuscation by also establishing the PUD-triangle to visualize the requirement preferences. 

\section{Overview of Dataset Obfuscation}
\label{sec_system}

This section introduces important definitions in each step of the obfuscation model and key performance used to evaluate the impacts on the resultant models in the crafted experiments.
% ; see Section~\ref{sec_experiments} for more details.

\begin{itemize}
    \item \textbf{Model Training}
\end{itemize}
% \subsection{Model Training}
Given a model $\mathcal{M}$ and a dataset $T=\{t_i,l_i\}$ with $t_i$ being the $i$-th sample in the dataset and $l_i$ being the label of the sample, the model weights $W$ can be trained with the training function $\mathcal F$.
\begin{equation}
    W=\mathcal F(T, \mathcal{M}).
\end{equation}
The trained  weights $W$ can be tested on a testing dataset $T_e$ and evaluated with the model prediction accuracy $\mathcal A(W)$.

\begin{itemize}
    \item \textbf{Data Obfuscation}
\end{itemize}
% \subsection{Data Obfuscation}
We consider a uniform obfuscation of training data where IID noises are added to the training data, denoted by $T'=\text{ }\mathcal O(T)$. The obfuscation function $\mathcal O(T)$ can be given by
\begin{equation}
\begin{split}    
&t_{ij}'=\text{ }t_{ij}+ \delta, \forall \text{ } i \text{ and } j,\ \delta \sim \mathcal N(0,\sigma^2);\\
&l_{i}'=\text{ }l_{i};
\end{split}
\end{equation}
where $t_{ij}$ is the $j$-th feature of $t_i$, and $t'_{ij}$ is the obfuscated version in $T'$. Labels, i.e., $l_{i}'$ and $l_{i}$, remain the same after data obfuscation.
$\delta \sim \mathcal N(0,\sigma^2)$ stands for the added Gaussian noise. The mean of the Gaussian noise is zero to ensure data utility. Here we consider all training data is obfuscated by setting $R=1$, where $R$ is the proportion of obfuscated data in the training dataset. 
 
% An example is given in Fig.~\ref{fig: experiment-add-noise}.
% The same training process can be applied to the obfuscated dataset. 
% We let $W'$ be the model weights trained from the obfuscated dataset $T'$, i.e., $W'=\mathcal F(T')$.

\begin{itemize}
    \item \textbf{Model Difference}
\end{itemize}
% \subsection{Model Difference}
We use the F-norm, i.e., $\|\cdot\|_{\text{F}}$, to measure the difference between two model weights, as given by
\begin{equation}
\begin{split}    
    \mathcal{D}(W_i,W_j)&=\|W_i-W_j\|_{\text{F}}=\sqrt {\sum_k|w_{ik}-w_{jk}|^{2}},
\end{split}
\end{equation}

where $w_{ik}$ and $w_{jk}$ are the $k$-th entries of $W_i$ and $W_j$, respectively.

\section{Experiments}
\label{sec_experiments}
In this section, we experimentally evaluate the impacts of adding random noise to training datasets on the resultant models. Experiments cover a range of Gaussian standard deviations and various IID and non-IID dataset settings.
\setcounter{subfigure}{0}
\begin{figure}[!ht]
\centering
\vspace{-0.35cm}
\subfigcapskip=-4pt
\subfigbottomskip=-4pt
\subfigure[$\sigma=0$]{
\label{fig_iotcoin}
\includegraphics[width=.25\columnwidth]{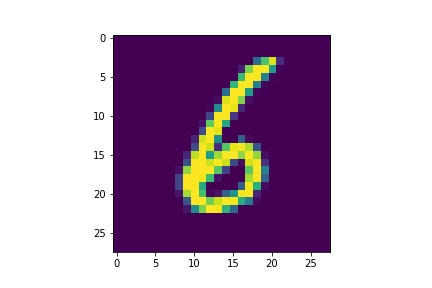}
}
\subfigure[$\sigma=0.4$]{
\label{fig_iotracking}
\includegraphics[width=.25\columnwidth]{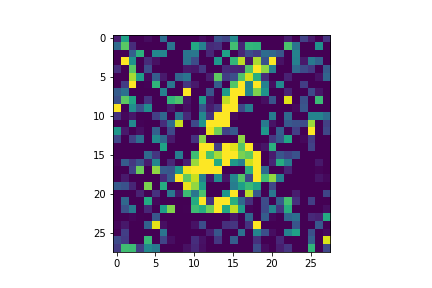}
}
\subfigure[$\sigma=0.8$]{
\label{fig_iottemp}
\includegraphics[width=.25\columnwidth]{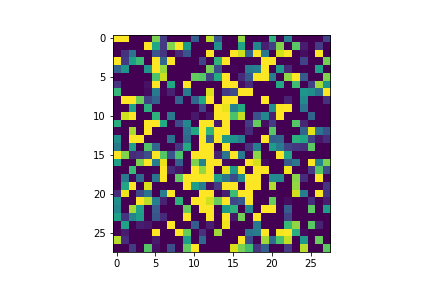}
}

\subfigure[$\sigma=0$]{
\label{fig_iotcoin2}
\includegraphics[width=.25\columnwidth]{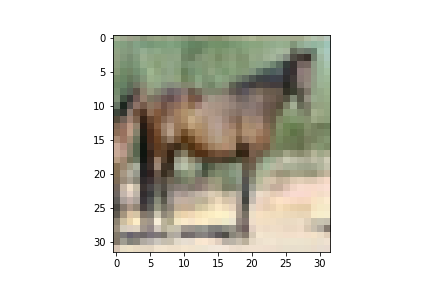}
}
\subfigure[$\sigma=0.4$]{
\label{fig_iotracking2}
\includegraphics[width=.25\columnwidth]{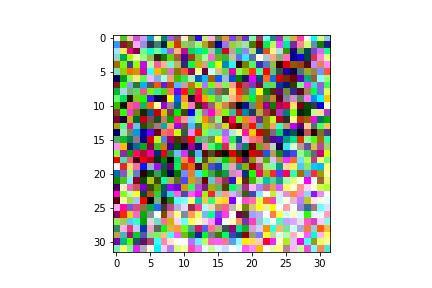}
}
\subfigure[$\sigma=0.8$]{
\label{fig_iottemp2}
\includegraphics[width=.25\columnwidth]{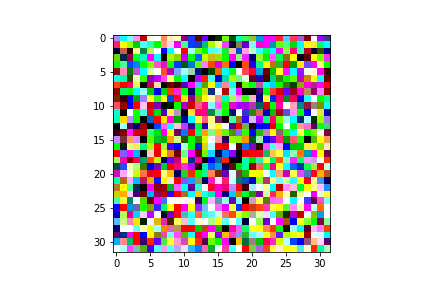}
}

\caption{Obfuscated MNIST (a, b, c) and CIFAR-10 (d, e, f).
% Adding Gaussian noise to the MNIST ( a, b, and c) and CIFAR-10 ( d, e, and f ) datasets with $R=1,\mu=0$, respectively.
}
\label{fig: experiment-add-noise}
	\vspace{-1em}  % 调整与下文的间距

\end{figure}

\subsection{Experiment Framework}\label{subsec: experiement settings}

\subsubsection{System Settings}

% \begin{itemize}
%     \item 2x Intel(R) Xeon(R) Gold 6138 CPU @ 2.00GHz 
%     \item 256GB RAM
%     \item 16TB of Disk Space
%     \item 8x NVIDIA A100 40GB PCIe (GPU)
% \end{itemize}

% \subsubsection{Software settings}
 We train classification models using popular MNIST (10 labels and 60,000 training examples) and CIFAR-10 (10 labels and 50,000 training examples) datasets. Each of the datasets is randomly split into a global training dataset $T_t$ with 90\% data and a testing dataset $T_e$ with the remaining 10\% data. The classification model is a standard Convolutional Neural Network (CNN) with six convolution layers and three max-pooling layers, named as $\mathcal{M}_{cnn}$. An initial model with random weights is used across all experiments. $\mathcal{M}_{cnn}$ is trained with 180 epochs and a learning rate of 0.0001 on both datasets. 
 The batch sizes for the training on MNIST and CIFAR-10 are 128 and 64, respectively.
 
%  Training configurations are given in Table~\ref{tab:my_label}. 

% \renewcommand\arraystretch{0.9}
% \begin{table}[!htb]
%     \caption{Training configuration of the MNIST and CIFAR-10 datasets.}
%     \label{tab:my_label}
%     \centering
% \begin{tabular}{|l|l|l|}
% \hline
% \textbf{Dataset}    & \textbf{MNIST}         & \textbf{CIFAR-10}  \\ \hline
% Classes    & 10         & 10  \\ \hline
% Dataset size& 60,000 & 50,000 \\ \hline
% Model      & $\mathcal{M}_{cnn}$ & $\mathcal{M}_{cnn}$ \\ \hline
% Epoch      & 180            & 180 \\ \hline
% Batch size & 128 & 64 \\ \hline
% Learning rate & 0.0001 &  0.0001 \\ \hline
% \end{tabular}
% \end{table}

We conduct the experiments upon Ubuntu 20.04.3 with TensorFlow 2.7.0 GPU kernel in Python 3.7. The training runs on a local machine with the following specifications:  2x Intel(R) Xeon(R) Gold 6138 CPU @ 2.00GHz, 256GB RAM, 16TB of Disk Space, and 8x NVIDIA A100 40GB.

% The elements in MNIST are images. Each image can described by a $d$-dimensional vector. The machine learning process here can be considered as a mapping, denoted by $\mathcal{X} \rightarrow \mathcal{Y}$. The input $\mathcal{X}$ is the features of training figures. $\mathcal{Y}$ includes a discrete set of classes. 
% Note that The fixed random seed is configured for all training processes to ensure reproducibility. Also note that this paper considers a 100\%-ratio of data obfuscation scheme ($R=1$) where all data samples are obfuscated by a zero-mean ($\mu=0$) random noise which is used to align with the data utility.

\subsubsection{Training Dataset Settings}
% We generate $n$ sub-datasets $T_{t}^i$ ($i \in \lbrace\lbrack 0,n )|\mathbb{Z}\rbrace$) from the global training dataset $T_{g}$ by IID and non-IID means. 
% The first dataset $T_{t}^0$, which is used as the control group, covers all labels and contains {\color{red}all} samples from $T_{g}$. 
We generate the training datasets from the global $T_t$ on IID and non-IID features, respectively, to show the impacts of training data on trained models. 
In the case of a non-IID dataset, we follow the non-IID data setting in~\cite{li2021federated} and consider Label-Skew-based and Quantity-Skew-based non-IID cases. We use ``S-$\mathcal{X}$-$\mathcal{Y}$-$\mathcal{Z}$'' to represent the sampling settings of training datasets.
% in the following Section~\ref{subsec: experiement results}. 
% Note that S-$1$-$1$-$1$ is equivalent to $T_{t}^0$.
\begin{itemize}
    \item Label degree $\mathcal{X}$: The dataset covers $\lfloor\mathcal{CX}\rfloor$ of the total $C$ labels, where $C=10$ for both datasets in this paper.
     
    % Each $T_{tr}^i$ contains a set of labels that have a ratio of $\mathcal{X}$ out from the completed label.
\end{itemize}

\begin{itemize}
    \item Label overlapping ratio $\mathcal{Y}$: 
    The dataset and its counterpart cover the same $\lfloor C\mathcal{X}\mathcal{Y}\rfloor$ labels.
    % the overlapping rate of labels between training datasets $T_{tr}^i$ and $T_{tr}^{i'}$ with $i\neq i'$ owned by different parties. Each $T_{tr}^i$ contains data samples evenly with overlapping labels at a ratio of $\mathcal{Y}$.
\end{itemize}

\begin{itemize}
    \item Sampling ratio $\mathcal{Z}$: 
    The dataset samples the entries of $T_t$ with ratio $\mathcal Z$ for each of the covered labels. The dataset contains $\mathcal{X}\mathcal{Z}|T_t|$ entries, where $|\cdot|$ gives cardinality.
    % the shrinkage $\mathcal{Z}$ of the data size of a training dataset. Each $T_{tr}^i$ contains identical labels but different amounts of data. 
\end{itemize}

%%%%%%%%%%%%%%%%%%%%%
\subsection{Experiment Results and Evaluations}\label{subsec: experiement results}

\begin{figure}
    \centering
    \subfigure[F-norm]{
    \begin{minipage}[t]{0.4\textwidth}
    \centering
    \includegraphics[width=2.2in]{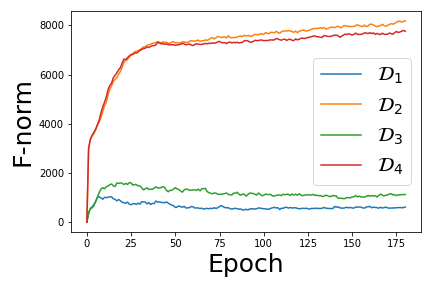}
    \end{minipage}
    \label{fig:experiment-training-norm}
    }
    \subfigure[Accuracy]{
    \begin{minipage}[t]{0.4\textwidth}
    \centering
    \includegraphics[width=2.2in]{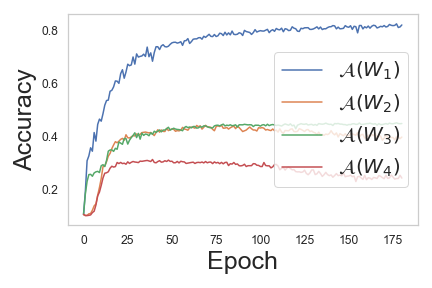}
    \end{minipage}
    \label{fig:experiment-training-accuracy}
    }
\caption{The dynamic F-norm and accuracy over epochs.}
\label{fig: experiment-epoch}
	\vspace{-1em}  % 调整与下文的间距

\end{figure}

A first look at the data obfuscation over MNIST and CIFAR-10 by adding the random Gaussian noises is shown in Fig.~\ref{fig: experiment-add-noise}.
% \vspace{0em}

We show the dynamics of model difference and model accuracy during training with CIFAR-10 in Fig.~\ref{fig: experiment-epoch} with setup in Table~\ref{tab_fig2}.
A reference dataset $T_{r}$ is sampled from a global $T_t$ with setting S-1-1-0.5, i.e., covering all labels and containing half of the global data. $T_{r}$ is also used as $T_{1}$, i.e., $T_{1}=T_{r}$, and obfuscated with $\sigma=1$ for $T_{o,2}$, i.e., $T_{o,2}=\mathcal O(T_{r}), \sigma=1$.
Dataset $T_{3}$ is sampled from  $T_t$ with setting S-0.5-1-1, i.e., covering half labels and containing all entries with the covered labels. $T_{o,4}$ is obfuscated $T_{3}$, i.e., $T_{o,4}=\mathcal O(T_{3})$, $\sigma=1$. 
All datasets contain $0.5|T_t|$ entries.
$W_r$ and $W_{i}$ can then be obtained by training the initial model on datasets.  
Next, the model distance can be computed via $\mathcal D_{i}=\mathcal D(W_r, W_{i})$.
Models are evaluated on $T_e$ for accuracy $\mathcal A(W)$. Results in Fig.~\ref{fig: experiment-epoch} show the convergence of both F-norm and accuracy over all (non-)IID and (non-)obfuscated settings. Note that the number of epochs used in all the experiments in this paper complies with that of Fig.~\ref{fig: experiment-epoch}.

%\vspace{0em}

We proceed to conduct four experiments over the MNIST and CIFAR-10 datasets from the following perspectives.

\begin{figure*}
    \centering
    \subfigure[\textbf{Exp-1} over MNIST]{
    \begin{minipage}[t]{0.23\textwidth}
    \centering
    \includegraphics[width=1.4in]{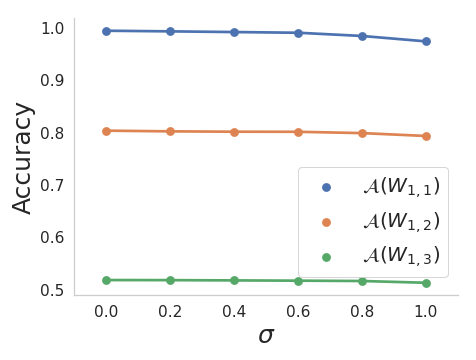}
    \end{minipage}
    \label{fig:experiment-acc-mnist-degree}
    }
    \subfigure[\textbf{Exp-1} over CIFAR-10]{
    \begin{minipage}[t]{0.23\textwidth}
    \centering
    \includegraphics[width=1.4in]{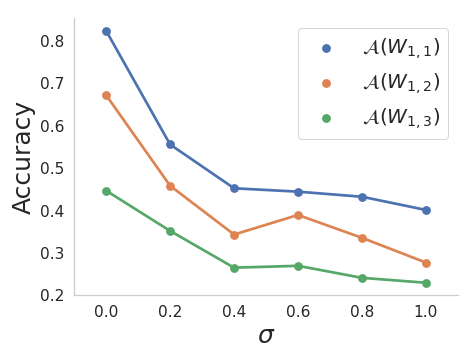}
    \end{minipage}
    \label{fig:experiment-acc-CIFAR-10-degree}
    }
    \subfigure[\textbf{Exp-2} over MNIST]{
    \begin{minipage}[t]{0.23\textwidth}
    \centering
    \includegraphics[width=1.4in]{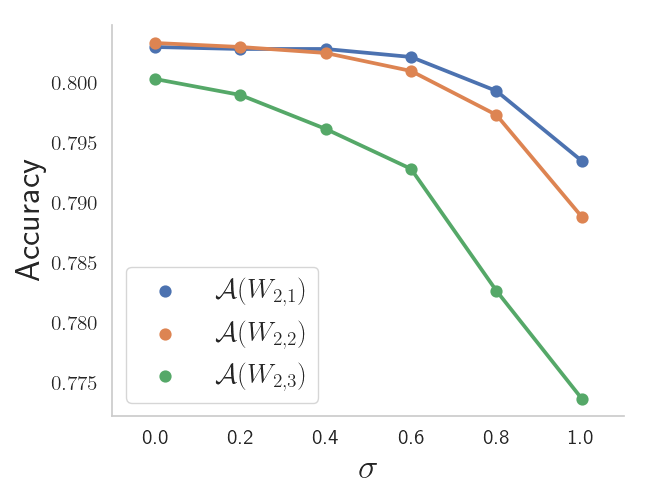}
    \end{minipage}
    \label{fig:experiment-acc-mnist-quantity}
    }
    \subfigure[\textbf{Exp-2} over CIFAR-10]{
    \begin{minipage}[t]{0.23\textwidth}
    \centering
    \includegraphics[width=1.4in]{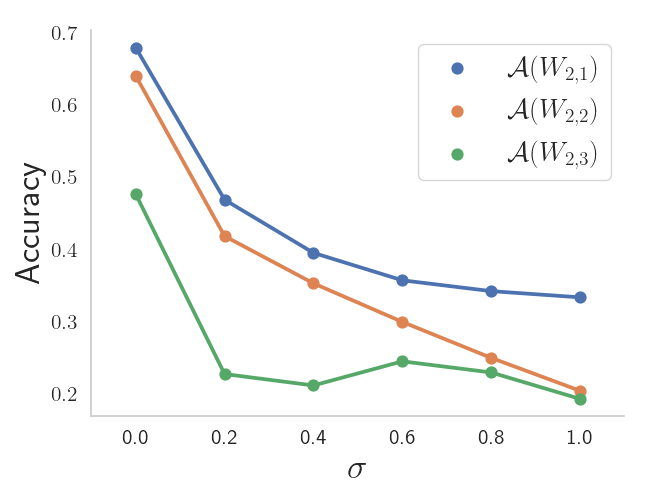}
    \end{minipage}
    \label{fig:experiment-acc-CIFAR-10-quantity}
    }
    
\caption{Model prediction accuracy over the testing dataset $T_e$ along with the increase of noise level $\sigma$ under different label degrees $\mathcal{X}$ and sampling ratios $\mathcal{Z}$ in MNIST and CIFAR-10 datasets, respectively. The $x$-axis is $\sigma$, and the $y$-axis gives the accuracy.}
\label{fig: experiment-accuracy}
\vspace{-1em}
\end{figure*}

\textbf{Exp-1 ($\mathcal A(W)$ vs. $\sigma$ with different label degrees)} - Figs.~\ref{fig:experiment-acc-mnist-degree} and~\ref{fig:experiment-acc-CIFAR-10-degree} with setup in Table~\ref{tab_exp1}. 
The model weight $W_{1,1}$ is trained on the obfuscated dataset $T_{o,1,1}=\mathcal{O}(T_{1,1})$ where $T_{1,1}$ is sampled from $T_t$ with setting S-1-1-0.5, and then evaluated on $T_e$ for $\mathcal{A}(W_{1,1})$. $\mathcal{A}(W_{1,2})$ and $\mathcal{A}(W_{1,3})$ are obtained in the same way with settings S-0.8-1-0.625 and S-0.5-1-1, respectively. All the datasets contain $0.5|T_t|$ entries.

\begin{table}[t]
    \centering
    \caption{Experiment setup for Fig.~\ref{fig: experiment-epoch}}
\resizebox{0.8\textwidth}{!}{%
\begin{tabular}{lll}
\hline
\multicolumn{1}{|l|}{\textbf{Dataset}}         & \multicolumn{1}{l|}{\textbf{Composition}}                                                                                                 & \multicolumn{1}{l|}{\textbf{Size}}                                                \\ \hline
\multicolumn{1}{|l|}{$T_{t}$}              & \multicolumn{1}{l|}{90\% of CIFAR-10 dataset}    & \multicolumn{1}{l|}{45,000}                                                         \\ \hline
\multicolumn{1}{|l|}{$T_{e}$}              & \multicolumn{1}{l|}{Rest 10\% CIFAR-10 dataset}    & \multicolumn{1}{l|}{5,000}                                     \\ \hline
\multicolumn{1}{|l|}{$T_{1}$ ($T_r$)}              & \multicolumn{1}{l|}{Sampled from $T_t$, with $\mathcal X=1$, $\mathcal Y=1$,   $\mathcal Z=0.5$}    & \multicolumn{1}{l|}{0.5$|T_t|$}                                                         \\ \hline
\multicolumn{1}{|l|}{$T_{o,2}$}                     & \multicolumn{1}{l|}{$\mathcal O(T_{1})$ with $\sigma=1$}                                                                           & \multicolumn{1}{l|}{$0.5|T_t|$}                                                         \\ \hline
\multicolumn{1}{|l|}{$T_{3}$}                     & \multicolumn{1}{l|}{Sampled from $T_t$ with $\mathcal X=0.5$, $\mathcal Y=1$, $\mathcal Z=1$}     & \multicolumn{1}{l|}{$0.5|T_t|$}                                                         \\ \hline
\multicolumn{1}{|l|}{$T_{o,4}$}                     & \multicolumn{1}{l|}{$\mathcal O(T_{3})$ with $\sigma=1$}                                                                           & \multicolumn{1}{l|}{$0.5|T_t|$}                                                          \\ \hline
\multicolumn{1}{|l|}{\textbf{Figures}}         & \multicolumn{1}{l|}{\textbf{Operations}}                                                                                                             & \multicolumn{1}{l|}{\textbf{Outputs}}                                             \\ \hline
\multicolumn{1}{|l|}{\multirow{2}{*}{Fig.~\ref{fig:experiment-training-norm}}} & \multicolumn{1}{l|}{Train $\mathcal{M}_{cnn}$ on $T_r$, $T_{1}$, $T_{o,2}$, $T_{3}$, $T_{o,4}$} & \multicolumn{1}{l|}{$W_r$, $W_{1}$, $W_{2}$, $W_{3}$, $W_{4}$}       \\\cline{2-3}  
\multicolumn{1}{|l|}{}                         & \multicolumn{1}{l|}{Calculate $\mathcal D(\cdot)$ between $W_r$ and {[}$W_{1}$, $W_{2}$, $W_{3}$, $W_{4}${]}}                            & \multicolumn{1}{l|}{$D_1$, $D_2$, $D_3$, $D_4$}                                       \\ \hline
\multicolumn{1}{|l|}{Fig.~\ref{fig:experiment-training-accuracy}}                  & \multicolumn{1}{l|}{Evaluate model accuracy over $T_e$}                             & \multicolumn{1}{l|}{$\mathcal A(W_{1})$, $\mathcal A(W_{2})$, $\mathcal A(W_{3})$, $\mathcal A(W_{4})$} \\ \hline
\end{tabular}
}
\label{tab_fig2}
\end{table}

% 对应代码 exp 17. Fig 1 Accuracy
\begin{table}[ht]
    \centering
    \caption{Experiment setup for Exp-1}
\resizebox{0.8\textwidth}{!}{%
\begin{tabular}{lll}
\hline
\multicolumn{1}{|l|}{\textbf{Dataset}}         & \multicolumn{1}{l|}{\textbf{Composition}}                                                                                                 & \multicolumn{1}{l|}{\textbf{Size}}                                                \\ \hline
\multicolumn{1}{|l|}{$T_{t}$}              & \multicolumn{1}{l|}{90\% of MNIST/CIFAR-10 dataset}    & \multicolumn{1}{l|}{54,000/45,000}                                                         \\ \hline
\multicolumn{1}{|l|}{$T_{e}$}              & \multicolumn{1}{l|}{Rest 10\% MNIST/CIFAR-10 dataset}    & \multicolumn{1}{l|}{6,000/5,000}                                                         \\ \hline
\multicolumn{1}{|l|}{$T_{1,1}$}              & \multicolumn{1}{l|}{Sampled from $T_t$, with $\mathcal X=1$, $\mathcal Y=1$,   $\mathcal Z=0.5$}    & \multicolumn{1}{l|}{0.5$|T_t|$}                                                         \\ \hline
\multicolumn{1}{|l|}{$T_{1,2}$}                     & \multicolumn{1}{l|}{Sampled from $T_t$ with $\mathcal X=0.8$, $\mathcal Y=1$, $\mathcal Z=0.625$}     & \multicolumn{1}{l|}{$0.5|T_t|$}            \\ \hline
\multicolumn{1}{|l|}{$T_{1,3}$}                     & \multicolumn{1}{l|}{Sampled from $T_t$ with $\mathcal X=0.5$, $\mathcal Y=1$, $\mathcal Z=1$}     & \multicolumn{1}{l|}{$0.5|T_t|$}            
\\ \hline
\multicolumn{1}{|l|}{\textbf{Figures}}         & \multicolumn{1}{l|}{\textbf{Operations}}                                                                                                             & \multicolumn{1}{l|}{\textbf{Outputs}}                                             \\ \hline
\multicolumn{1}{|l|}{\multirow{3}{*}{\begin{tabular}[c]{@{}l@{}}Fig.~\ref{fig:experiment-acc-mnist-degree}\\ MNIST\end{tabular}}} & \multicolumn{1}{l|}{Execute $\mathcal O(\cdot)$ on $T_{1,1}$, $T_{1,2}$, and $T_{1,3}$} & \multicolumn{1}{l|}{$T_{o,1,1}$, $T_{o,1,2}$, $T_{o,1,3}$}       \\\cline{2-3}
\multicolumn{1}{|l|}{}                         &\multicolumn{1}{l|}{Train $\mathcal{M}_{cnn}$ on $T_{o,1,1}$, $T_{o,1,2}$, $T_{o,1,3}$} & \multicolumn{1}{l|}{$W_{1,1}$, $W_{1,2}$, and $W_{1,3}$}       \\\cline{2-3}
\multicolumn{1}{|l|}{}                         & \multicolumn{1}{l|}{Evaluate model accuracy over $T_e$ }& \multicolumn{1}{l|}{$\mathcal A(W_{1,1})$, $\mathcal A(W_{1,2})$, $\mathcal A(W_{1,3})$}                                       \\ \hline
\multicolumn{1}{|l|}{\begin{tabular}[c]{@{}l@{}}Fig.~\ref{fig:experiment-acc-CIFAR-10-degree}\\ CIFAR-10\end{tabular}}                  & \multicolumn{1}{l|}{Same as above}                             & \multicolumn{1}{l|}{Same as above} \\ \hline
\end{tabular}
}
\label{tab_exp1}
\end{table}

\textbf{Exp-2 ($\mathcal A(W)$ vs. $\sigma$ with different dataset sizes)} - Figs.~\ref{fig:experiment-acc-mnist-quantity} and~\ref{fig:experiment-acc-CIFAR-10-quantity} with setup in Table~\ref{tab_exp2}. 
The model weight $W_{2,1}$ is trained on the obfuscated dataset $T_{o,2,1}=\mathcal{O}(T_{2,1})$ where $T_{2,1}$ is sampled from $T_t$ with setting S-0.8-1-1, and then evaluated on $T_e$ for $\mathcal{A}(W_{2,1})$. 
$\mathcal{A}(W_{2,2})$ and $\mathcal{A}(W_{2,3})$ are obtained in the same way with settings S-0.8-1-0.5 and S-0.8-1-0.1, respectively. $T_{2,1}$, $T_{2,2}$ and $T_{2,3}$ cover the same eight labels and contain $0.8|T_t|$, $0.4|T_t|$, and $0.08|T_t|$ entries, respectively.

% 对应代码 exp 17. Fig 2 Accuracy
\begin{table}[ht]
    \centering
    \caption{Experiment setup for Exp-2}
\resizebox{0.8\textwidth}{!}{%
\begin{tabular}{lll}
\hline
\multicolumn{1}{|l|}{\textbf{Dataset}}         & \multicolumn{1}{l|}{\textbf{Composition}}                                                                                                 & \multicolumn{1}{l|}{\textbf{Size}}                                                \\ \hline
\multicolumn{1}{|l|}{$T_{t}$}              & \multicolumn{1}{l|}{90\% of MNIST/CIFAR-10 dataset}    & \multicolumn{1}{l|}{54,000/45,000}                                                         \\ \hline
\multicolumn{1}{|l|}{$T_{e}$}              & \multicolumn{1}{l|}{Rest 10\% MNIST/CIFAR-10 dataset}    & \multicolumn{1}{l|}{6,000/5,000}                                                         \\ \hline
\multicolumn{1}{|l|}{$T_{2,1}$}              & \multicolumn{1}{l|}{Sampled from $T_t$, with $\mathcal X=0.8$, $\mathcal Y=1$,   $\mathcal Z=1$}    & \multicolumn{1}{l|}{0.8$|T_t|$}                                                         \\ \hline
\multicolumn{1}{|l|}{$T_{2,2}$}                     & \multicolumn{1}{l|}{Random 50\% data of $T_{2,1}$ }     & \multicolumn{1}{l|}{$0.4|T_t|$}            \\ \hline
\multicolumn{1}{|l|}{$T_{2,3}$}                     & \multicolumn{1}{l|}{Random 10\% data of $T_{2,1}$}     & \multicolumn{1}{l|}{$0.08|T_t|$}            
\\ \hline
\multicolumn{1}{|l|}{\textbf{Figures}}         & \multicolumn{1}{l|}{\textbf{Operations}}                                                                                                             & \multicolumn{1}{l|}{\textbf{Outputs}}                                             \\ \hline
\multicolumn{1}{|l|}{\multirow{3}{*}{\begin{tabular}[c]{@{}l@{}}Fig.~\ref{fig:experiment-acc-mnist-quantity}\\ MNIST\end{tabular}}} & \multicolumn{1}{l|}{Execute $\mathcal O(\cdot)$ on $T_{2,1}$, $T_{2,2}$, and $T_{2,3}$} & \multicolumn{1}{l|}{$T_{o,2,1}$, $T_{o,2,2}$, $T_{o,2,3}$}       \\\cline{2-3}
\multicolumn{1}{|l|}{}                         &\multicolumn{1}{l|}{Train $\mathcal{M}_{cnn}$ on $T_{o,2,1}$, $T_{o,2,2}$, $T_{o,2,3}$} & \multicolumn{1}{l|}{$W_{2,1}$, $W_{2,2}$, and $W_{2,3}$}       \\\cline{2-3}
\multicolumn{1}{|l|}{}                         & \multicolumn{1}{l|}{Evaluate model accuracy over $T_e$ }& \multicolumn{1}{l|}{$\mathcal A(W_{2,1})$, $\mathcal A(W_{2,2})$, $\mathcal A(W_{2,3})$}                                       \\ \hline
\multicolumn{1}{|l|}{\begin{tabular}[c]{@{}l@{}}Fig.~\ref{fig:experiment-acc-CIFAR-10-quantity}\\ CIFAR-10\end{tabular}}                  & \multicolumn{1}{l|}{Same as above}                             & \multicolumn{1}{l|}{Same as above} \\ \hline
\end{tabular}
}
\label{tab_exp2}
\end{table}

\begin{figure*}
    \centering
    \subfigure[\textbf{Exp-3} over MNIST]{
    \begin{minipage}[t]{0.23\textwidth}
    \centering
    \includegraphics[width=1.4in]{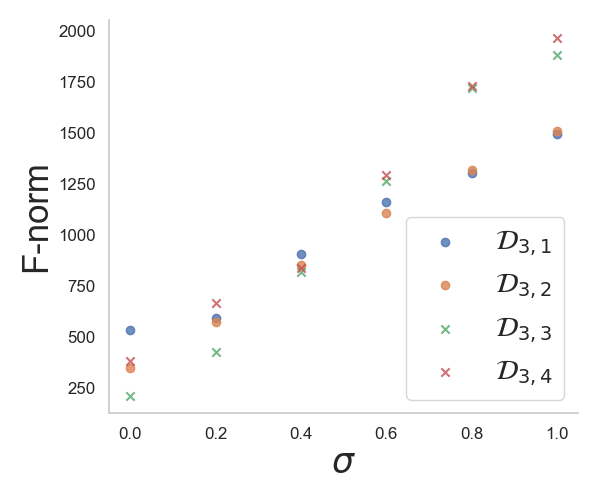}
    \end{minipage}
    \label{fig:experiment-norm-mnist-label}
    }
    \subfigure[\textbf{Exp-3} over CIFAR-10]{
    \begin{minipage}[t]{0.23\textwidth}
    \centering
    \includegraphics[width=1.4in]{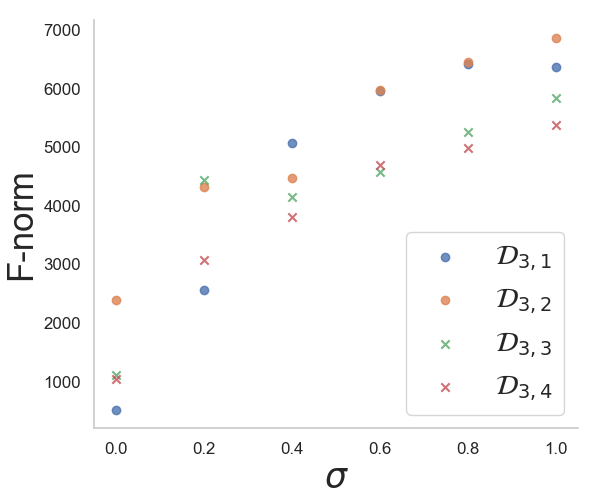}
    \end{minipage}
    \label{fig:experiment-norm-CIFAR-10-label}
    }
    \subfigure[\textbf{Exp-4} over MNIST]{
    \begin{minipage}[t]{0.23\textwidth}
    \centering
    \includegraphics[width=1.4in]{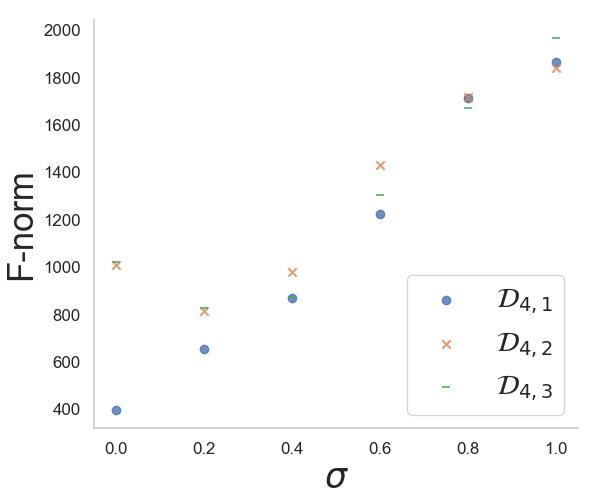}
    \end{minipage}
    \label{fig:experiment-norm-mnist-overlap}
    }
    \subfigure[\textbf{Exp-4} over CIFAR-10]{
    \begin{minipage}[t]{0.23\textwidth}
    \centering
    \includegraphics[width=1.4in]{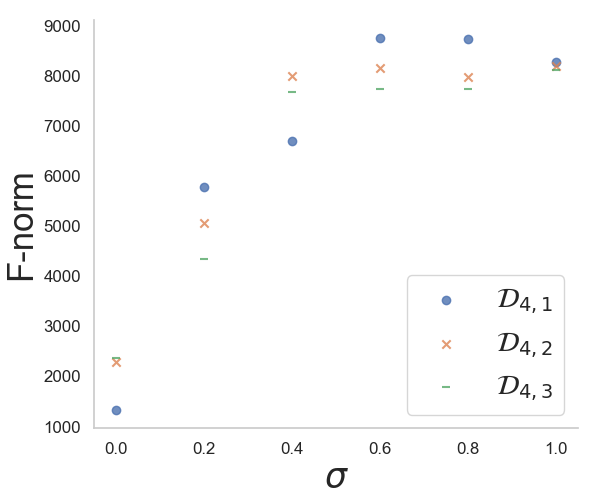}
    \end{minipage}
    \label{fig:experiment-norm-CIFAR-10-overlap}
    }
    % \subfigure[\textbf{Exp-5} over MNIST]{
    % \begin{minipage}[t]{0.225\textwidth}
    % \centering
    % \includegraphics[width=1.7in]{graphs/exp-results/Fig-5a.png}
    % \end{minipage}
    % \label{fig:experiment-norm-mnist-label-5}
    % }
    % \subfigure[\textbf{Exp-5} over CIFAR-10]{
    % \begin{minipage}[t]{0.225\textwidth}
    % \centering
    % \includegraphics[width=1.7in]{graphs/exp-results/Fig-5b.png}
    % \end{minipage}
    % \label{fig:experiment-norm-CIFAR-10-label-5}
    % }
    
\caption{F-norm-based model difference between models trained from reference datasets without noise and obfuscated datasets along with the increase of noise level $\sigma$. The $x$-axis is $\sigma$ for Gaussian noise, and the $y$-axis gives the F-norm values.}
\label{fig: experiment-fnorm}
\vspace{-1em}  % 调整与下文的间距
\end{figure*}

\textbf{Exp-3 ($\mathcal{D}(W_{r},W_{o})$ vs. $\sigma$ with different label degrees)} - Figs. \ref{fig:experiment-norm-mnist-label} and~\ref{fig:experiment-norm-CIFAR-10-label} with setup in Table~\ref{tab_exp3}. $W_{r}$ is obtained by training reference datasets $T_r$ without any noise, while $W_{o}$ is from the obfuscated datasets $T_o$. Three sampling settings are designed to capture the impact of noise on models from IID datasets under different label degrees.
Setting S-1-1-0.25 is employed for $\mathcal{D}_{3,1}$ and $\mathcal{D}_{3,2}$.
$\mathcal{D}_{3,1}=\mathcal{D}(W_{r,3,1},W_{o,3,1})$, where $W_{r,3,1}$ is from dataset $T_{r,3,1}$ sampled from $T_t$ with setting S-1-1-0.25. The reference dataset is also used for the obfuscated dataset for $W_{o,3,1}$, i.e., $T_{o,3,1}$=$O(T_{r,3,1})$. 
In terms of $\mathcal{D}_{3,2}=\mathcal D(W_{r,3,1},W_{o,3,2})$, $T_{o,3,2}=O(T_{3,2})$ where $T_{3,2}$ is sampled from $T_t$ following the same setting S-1-1-0.25. All the datasets for $\mathcal{D}_{3,1}$ and $\mathcal{D}_{3,2}$ cover all the ten labels. 
$\mathcal{D}_{3,3}$ and $\mathcal{D}_{3,4}$ are obtained with setting S-0.5-1-0.5, in which all datasets cover the first five labels. All datasets in Exp-3 contain $0.25|T_t|$ entries.

% 对应代码 exp 17. Fig 3-1~3-3 F-norm，不过为了简化实验，Fig 3-2的内容被删除了，没有在图上体现

\begin{table}[ht]
    \centering
    \caption{Experiment setup for Exp-3}
\resizebox{0.8\textwidth}{!}{%
\begin{tabular}{lll}
\hline
\multicolumn{1}{|l|}{\textbf{Dataset}}         & \multicolumn{1}{l|}{\textbf{Composition}}                                                                                                 & \multicolumn{1}{l|}{\textbf{Size}}                                                \\ \hline
\multicolumn{1}{|l|}{$T_{t}$}              & \multicolumn{1}{l|}{90\% of MNIST/CIFAR-10 dataset}    & \multicolumn{1}{l|}{54,000/45,000}                                                         \\ \hline
\multicolumn{1}{|l|}{$T_{e}$}              & \multicolumn{1}{l|}{Rest 10\% MNIST/CIFAR-10 dataset}    & \multicolumn{1}{l|}{6,000/5,000}                                                         \\ \hline
\multicolumn{1}{|l|}{$T_{3,1}$ ($T_{r,3,1}$)}              & \multicolumn{1}{l|}{Sampled from $T_t$, with $\mathcal X=1$, $\mathcal Y=1$,   $\mathcal Z=0.25$}    & \multicolumn{1}{l|}{$0.25|T_t|$}                                                         \\ \hline
\multicolumn{1}{|l|}{$T_{3,2}$}                     & \multicolumn{1}{l|}{Sampled from $T_t$, with the same setting and labels of $T_{3,1}$}     & \multicolumn{1}{l|}{$0.25|T_t|$}            \\ \hline
\multicolumn{1}{|l|}{$T_{3,3}$ ($T_{r,3,2}$)}              & \multicolumn{1}{l|}{Sampled from $T_t$, with $\mathcal X=0.5$, $\mathcal Y=1$,   $\mathcal Z=0.5$}    & \multicolumn{1}{l|}{$0.25|T_t|$}                                                         \\ \hline
\multicolumn{1}{|l|}{$T_{3,4}$}                     & \multicolumn{1}{l|}{Sampled from $T_t$, with the same setting and labels of $T_{3,3}$}     & \multicolumn{1}{l|}{$0.25|T_t|$}            \\ \hline
\multicolumn{1}{|l|}{\textbf{Figures}}         & \multicolumn{1}{l|}{\textbf{Operations}}                                                                                                             & \multicolumn{1}{l|}{\textbf{Outputs}}                                             \\ \hline
\multicolumn{1}{|l|}{\multirow{4}{*}{\begin{tabular}[c]{@{}l@{}}Fig.~\ref{fig:experiment-norm-mnist-label}\\ MNIST\end{tabular}}} & \multicolumn{1}{l|}{Execute $\mathcal O(\cdot)$ on $T_{3,1}$, $T_{3,2}$, $T_{3,3}$, and $T_{3,4}$ with a varying $\sigma$} & \multicolumn{1}{l|}{$T_{o,3,1}$, $T_{o,3,2}$, $T_{o,3,3}$, $T_{o,3,4}$}       \\\cline{2-3}
\multicolumn{1}{|l|}{}                         &\multicolumn{1}{l|}{Train $\mathcal{M}_{cnn}$ on $T_{r,3,1}$, $T_{r,3,2}$, $T_{o,3,1}$, $T_{o,3,2}$, $T_{o,3,3}$, and $T_{o,3,4}$} & \multicolumn{1}{l|}{$W_{r,3,1}$, $W_{r,3,2}$, $W_{o,3,1}$, $W_{o,3,2}$, $W_{o,3,3}$, $W_{o,3,4}$}       \\\cline{2-3}
\multicolumn{1}{|l|}{}                         & \multicolumn{1}{l|}{Calculate $\mathcal D(\cdot)$ between $W_{r,3,1}$ and [$W_{o,3,1}$, $W_{o,3,2}$]}& \multicolumn{1}{l|}{$\mathcal D_{3,1}$, $\mathcal D_{3,2}$}  \\\cline{2-3}
\multicolumn{1}{|l|}{}                         & \multicolumn{1}{l|}{Calculate $\mathcal D(\cdot)$ between $W_{r,3,2}$ and [$W_{o,3,3}$, $W_{o,3,4}$]}& \multicolumn{1}{l|}{$\mathcal D_{3,3}$, $\mathcal D_{3,4}$}                                       \\ \hline
\multicolumn{1}{|l|}{\begin{tabular}[c]{@{}l@{}}Fig.~\ref{fig:experiment-norm-CIFAR-10-label}\\ CIFAR-10\end{tabular}}                  & \multicolumn{1}{l|}{Same as above}                             & \multicolumn{1}{l|}{Same as above} \\ \hline
\end{tabular}
}
\label{tab_exp3}
\end{table}

\textbf{Exp-4 ($\mathcal{D}(W_r,W_o)$ vs. $\sigma$ with different label overlapping ratios)} - Figs.~\ref{fig:experiment-norm-mnist-overlap} and~\ref{fig:experiment-norm-CIFAR-10-overlap} with setup in Table~\ref{tab_exp4}.
$\mathcal{D}_{4,1}=\mathcal{D}(W_{r,4,1},W_{o,4,1})$, where $W_{r,4,1}$ is trained from dataset $T_{r,4,1}$ that is sampled from $T_t$ with setting S-0.5-1-0.5 and covers labels 0, 1, 2, 3 and 4. The reference dataset is also employed for the obfuscated dataset $W_{o,4,1}$, i.e., $T_{o,4,1}$=$O(T_{r,4,1})$, and the reference dataset in Exp-4. In terms of $\mathcal{D}_{4,2}=\mathcal D(W_{r,4,1}, W_{o,4,2})$, $T_{o,4,2}$ is an obfuscated $T_{4,2}$ that is sampled from $T_t$ with setting S-0.5-0.5-0.5 and covers labels 3, 4, 5, 6 and 7. For $\mathcal{D}_{4,3}=\mathcal D(W_{r,4,1}, W_{o,4,3})$, $T_{o,4,3}$ is an obfuscated $T_{4,3}$ that is sampled from $T_t$ with setting S-0.5-0.1-0.5 and covers labels 4, 5, 6, 7 and 8. All datasets contain $0.25|T_t|$ entries.

% 对应代码 exp 17. Fig 4 F-norm

\begin{table}[ht]
    \centering
    \caption{Experiment setup for Exp-4}
\resizebox{0.8\textwidth}{!}{%
\begin{tabular}{lll}
\hline
\multicolumn{1}{|l|}{\textbf{Dataset}}         & \multicolumn{1}{l|}{\textbf{Composition}}                                                                                                 & \multicolumn{1}{l|}{\textbf{Size}}                                                \\ \hline
\multicolumn{1}{|l|}{$T_{t}$}              & \multicolumn{1}{l|}{90\% of MNIST/CIFAR-10 dataset}    & \multicolumn{1}{l|}{54,000/45,000}                                                         \\ \hline
\multicolumn{1}{|l|}{$T_{e}$}              & \multicolumn{1}{l|}{Rest 10\% MNIST/CIFAR-10 dataset}    & \multicolumn{1}{l|}{6,000/5,000}                                                         \\ \hline
\multicolumn{1}{|l|}{$T_{4,1}$ ($T_{r,4,1}$)}              & \multicolumn{1}{l|}{Sampled from $T_t$, with $\mathcal X=0.5$, $\mathcal Y=1$,   $\mathcal Z=0.5$, covering labels 0-4}    & \multicolumn{1}{l|}{$0.25|T_t|$}                                                         \\ \hline
\multicolumn{1}{|l|}{$T_{4,2}$}              & \multicolumn{1}{l|}{Sampled from $T_t$, with $\mathcal X=0.5$, $\mathcal Y=0.4$,   $\mathcal Z=0.5$, covering labels 3-7}    & \multicolumn{1}{l|}{$0.25|T_t|$}                                                         \\ \hline
\multicolumn{1}{|l|}{$T_{4,3}$}              & \multicolumn{1}{l|}{Sampled from $T_t$, with $\mathcal X=0.5$, $\mathcal Y=0.2$,   $\mathcal Z=0.5$, covering labels 4-8}    & \multicolumn{1}{l|}{$0.25|T_t|$}                                                         \\ \hline
\multicolumn{1}{|l|}{\textbf{Figures}}         & \multicolumn{1}{l|}{\textbf{Operations}}                                                                                                             & \multicolumn{1}{l|}{\textbf{Outputs}}                                             \\ \hline
\multicolumn{1}{|l|}{\multirow{3}{*}{\begin{tabular}[c]{@{}l@{}}Fig.~\ref{fig:experiment-norm-mnist-overlap}\\ MNIST\end{tabular}}} & \multicolumn{1}{l|}{Execute $\mathcal O(\cdot)$ on $T_{4,1}$, $T_{4,2}$, and $T_{4,3}$ with a varying $\sigma$} & \multicolumn{1}{l|}{$T_{o,4,1}$, $T_{o,4,2}$, $T_{o,4,3}$}       \\\cline{2-3}
\multicolumn{1}{|l|}{}                         &\multicolumn{1}{l|}{Train $\mathcal{M}_{cnn}$ on $T_{r,4,1}$, $T_{o,4,1}$, $T_{o,4,2}$, and $T_{o,4,3}$} & \multicolumn{1}{l|}{$W_{r,4,1}$, $W_{o,4,1}$, $W_{o,4,2}$, $W_{o,4,3}$}       \\\cline{2-3}
\multicolumn{1}{|l|}{}                         & \multicolumn{1}{l|}{Calculate $\mathcal D(\cdot)$ between $W_{r,4,1}$ and [$W_{o,4,1}$, $W_{o,4,2}$, $W_{o,4,3}$]}& \multicolumn{1}{l|}{$\mathcal D_{4,1}$, $\mathcal D_{4,2}$, $\mathcal D_{4,3}$}
\\ \hline
\multicolumn{1}{|l|}{\begin{tabular}[c]{@{}l@{}}Fig.~\ref{fig:experiment-norm-CIFAR-10-overlap}\\ CIFAR-10\end{tabular}}                  & \multicolumn{1}{l|}{Same as above}                             & \multicolumn{1}{l|}{Same as above} \\ \hline
\end{tabular}
}
\label{tab_exp4}
\end{table}

\begin{figure}
    \centering
    \subfigure[\textbf{Exp-5} over MNIST]{
    \begin{minipage}[t]{0.3\textwidth}
    \centering
    \includegraphics[width=1.4in]{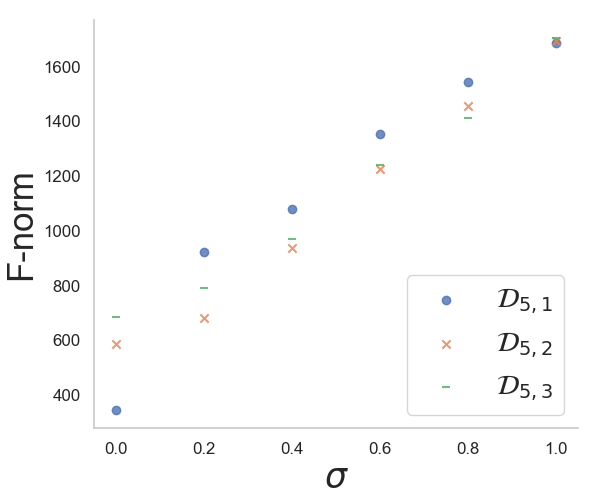}
    \end{minipage}
    \label{fig:experiment-norm-mnist-label-5}
    }
    \subfigure[\textbf{Exp-5} over CIFAR-10]{
    \begin{minipage}[t]{0.3\textwidth}
    \centering
    \includegraphics[width=1.4in]{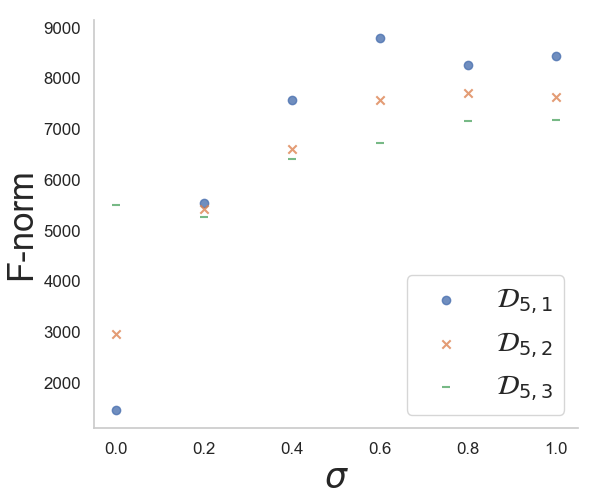}
    \end{minipage}
    \label{fig:experiment-norm-CIFAR-10-label-5}
    }
    
\caption{Model difference along with the increase of noise level $\sigma$. The $x$-axis is $\sigma$ for Gaussian noise, and the $y$-axis gives the F-norm values.}
\label{fig: experiment-fnorm-2}
	\vspace{-1em}  % 调整与下文的间距

\end{figure}

\textbf{Exp-5 ($\mathcal{D}(W_r,W_o)$ vs. $\sigma$ with different dataset sizes)} - Figs.~\ref{fig:experiment-norm-mnist-label-5} and~\ref{fig:experiment-norm-CIFAR-10-label-5} with setup in Table~\ref{tab_exp5}.
$\mathcal{D}_{5,1}=\mathcal{D}(W_{r,5,1},W_{o,5,1})$, where $W_{r,5,1}$ is trained on the dataset $T_{r,5,1}$ that is sampled from $T_t$ with setting S-1-1-0.5 and contains $0.5|T_t|$ entries. The reference dataset is also used for the obfuscated dataset $W_{o,5,1}$, i.e., $T_{o,5,1}$=$O(T_{r,5,1})$, and the other reference datasets in this experiment, i.e., $T_{r,5,3}=T_{r,5,2}=T_{r,5,1}$. In terms of $\mathcal{D}_{5,2}$, $T_{o,5,2}$ is an obfuscated $T_{5,2}$ that is sampled from $T_t$ with setting S-1-1-0.25 and contains $0.25|T_t|$ entries. For $\mathcal{D}_{5,3}$, $T_{o,5,3}$ is sampled from $T_t$ with setting S-1-1-0.05 and contains $0.05|T_t|$ entries.

% 对应代码 exp 17. Fig 5 F-norm

\begin{table}[ht]
    \centering
    \caption{Experiment setup for Exp-5}
\resizebox{0.8\textwidth}{!}{%
\begin{tabular}{lll}
\hline
\multicolumn{1}{|l|}{\textbf{Dataset}}         & \multicolumn{1}{l|}{\textbf{Composition}}                                                                                                 & \multicolumn{1}{l|}{\textbf{Size}}                                                \\ \hline
\multicolumn{1}{|l|}{$T_{t}$}              & \multicolumn{1}{l|}{90\% of MNIST/CIFAR-10 dataset}    & \multicolumn{1}{l|}{54,000/45,000}                                                         \\ \hline
\multicolumn{1}{|l|}{$T_{e}$}              & \multicolumn{1}{l|}{Rest 10\% MNIST/CIFAR-10 dataset}    & \multicolumn{1}{l|}{6,000/5,000}                                                         \\ \hline
\multicolumn{1}{|l|}{$T_{5,1}$ ($T_{r,5,1}$)}              & \multicolumn{1}{l|}{Sampled from $T_t$, with $\mathcal X=1$, $\mathcal Y=1$,   $\mathcal Z=0.5$}    & \multicolumn{1}{l|}{$0.5|T_t|$}                                                         \\ \hline
\multicolumn{1}{|l|}{$T_{5,2}$}              & \multicolumn{1}{l|}{Sampled from $T_t$, with $\mathcal X=1$, $\mathcal Y=1$,   $\mathcal Z=0.25$}    & \multicolumn{1}{l|}{$0.25|T_t|$}                                                         \\ \hline
\multicolumn{1}{|l|}{$T_{5,3}$}              & \multicolumn{1}{l|}{Sampled from $T_t$, with $\mathcal X=1$, $\mathcal Y=1$,   $\mathcal Z=0.05$}    & \multicolumn{1}{l|}{$0.05|T_t|$}                                                         \\ \hline
\multicolumn{1}{|l|}{\textbf{Figures}}         & \multicolumn{1}{l|}{\textbf{Operations}}                                                                                                             & \multicolumn{1}{l|}{\textbf{Outputs}}                                             \\ \hline
\multicolumn{1}{|l|}{\multirow{3}{*}{\begin{tabular}[c]{@{}l@{}}Fig.~\ref{fig:experiment-norm-mnist-label-5}\\ MNIST\end{tabular}}} & \multicolumn{1}{l|}{Execute $\mathcal O(\cdot)$ on $T_{5,1}$, $T_{5,2}$, and $T_{5,3}$ with a varying $\sigma$} & \multicolumn{1}{l|}{$T_{o,5,1}$, $T_{o,5,2}$, $T_{o,5,3}$}       \\\cline{2-3}
\multicolumn{1}{|l|}{}                         &\multicolumn{1}{l|}{Train $\mathcal{M}_{cnn}$ on $T_{r,5,1}$, $T_{o,5,1}$, $T_{o,5,2}$, and $T_{o,5,3}$} & \multicolumn{1}{l|}{$W_{r,5,1}$, $W_{o,5,1}$, $W_{o,5,2}$, $W_{o,5,3}$}       \\\cline{2-3}
\multicolumn{1}{|l|}{}                         & \multicolumn{1}{l|}{Calculate $\mathcal D(\cdot)$ between $W_{r,5,1}$ and [$W_{o,5,1}$, $W_{o,5,2}$, $W_{o,5,3}$]}& \multicolumn{1}{l|}{$\mathcal D_{5,1}$, $\mathcal D_{5,2}$, $\mathcal D_{5,3}$}
\\ \hline
\multicolumn{1}{|l|}{\begin{tabular}[c]{@{}l@{}}Fig.~\ref{fig:experiment-norm-CIFAR-10-label-5}\\ CIFAR-10\end{tabular}}                  & \multicolumn{1}{l|}{Same as above}                             & \multicolumn{1}{l|}{Same as above} \\ \hline
\end{tabular}
}
\label{tab_exp5}
\end{table}

Fig.~\ref{fig: experiment-accuracy} shows the decrease of the model accuracy with the increase of the noise level. The less diversified labels a training dataset contains, the lower accuracy the model can achieve over a fully-labeled testing dataset ($\mathcal{X}=1$), as shown in Figs.~\ref{fig:experiment-acc-mnist-degree} and~\ref{fig:experiment-acc-CIFAR-10-degree}. On the other hand, the models containing more data samples across covered labels can be more resistant to the adverse effect of obfuscation on the model accuracy, as shown in Figs.~\ref{fig:experiment-acc-mnist-quantity} and~\ref{fig:experiment-acc-CIFAR-10-quantity}. This indicates that the accumulation of knowledge from the obfuscated samples is useful for recognizing the correct patterns.

It can be found in Fig.~\ref{fig: experiment-fnorm} that the F-norm $\mathcal{D}(W_r,W_o)$ increases with the noise level.
% on a logarithmic scale
This reflects that the obfuscated samples lead the model into a different direction of gradient descent, thus weakening the training reproducibility.
In Figs.~\ref{fig:experiment-norm-mnist-label} and~\ref{fig:experiment-norm-CIFAR-10-label}, we conduct two types of comparisons: 
\begin{itemize}
\item A cross-comparison about the F-norm between points with different reference datasets ($\mathcal{D}_{3,1}$ and $\mathcal{D}_{3,2}$ vs. $\mathcal{D}_{3,3}$ and $\mathcal{D}_{3,4}$). Results show that, in terms of F-norm, $\mathcal{D}_{3,3}$ and $\mathcal{D}_{3,4}$ are greater than $\mathcal{D}_{3,1}$ and $\mathcal{D}_{3,2}$ when the noise is large over a relatively simpler training task MNIST but behaves in the opposite way over the CIFAR-10 dataset.
% {\color{red}This might be because the greater number of epochs required to converge the F-norm over a relatively more complicated training task enlarges the impacts of obfuscation.} 
Despite having more labels (larger $\mathcal{X}$) leads to higher model privacy as revealed in Figs.~\ref{fig:experiment-acc-mnist-quantity} and~\ref{fig:experiment-acc-CIFAR-10-quantity}, having a larger F-norm with a higher model accuracy implies the existence of multiple optima.

\item A comparison between points with the same reference datasets ($\mathcal{D}_{3,1}$ vs. $\mathcal{D}_{3,2}$ and $\mathcal{D}_{3,3}$ vs. $\mathcal{D}_{3,4}$). Results show that the gap of either comparison becomes indistinguishable in both two groups with the noise level increasing to $\sigma\approx 0.4$. In addition, the more diversified labels a dataset contains, the harder it can be discriminable from the other dataset, and the weaker the model can be resistant to the adverse effect of obfuscation on the model difference.
\end{itemize}

In Figs.~\ref{fig:experiment-norm-mnist-overlap} and~\ref{fig:experiment-norm-CIFAR-10-overlap}, the gaps between the reference points,  $\mathcal{D}_{4,1}$, and $\mathcal{D}_{4,2}$, and $\mathcal{D}_{4,3}$ at each given noise level are considered. The experiments on both MNIST and CIFAR-10 show that the gaps narrow down as the noise level increases. Given the same noise level, $\mathcal{D}_{4,3}$ tends to be more discriminable from $\mathcal{D}_{4,1}$ than $\mathcal{D}_{4,2}$ due to the lower label overlapping ratios $\mathcal{Y}$ between $\mathcal{D}_{4,1}$ and $\mathcal{D}_{4,3}$. Also, having such a lower $\mathcal{Y}$ can be more resistant to the impacts of obfuscation on the model difference, given that $\mathcal{X}$s are identical.

Figs.~\ref{fig:experiment-norm-mnist-label-5} and~\ref{fig:experiment-norm-CIFAR-10-label-5} are learned in concert with Figs.~\ref{fig:experiment-acc-mnist-quantity} and~\ref{fig:experiment-acc-CIFAR-10-quantity} in terms of the sampling ratio $\mathcal{Z}$. Given a certain number of epochs and identical $\mathcal{X}$s and $\mathcal{Y}$s, a smaller $\mathcal{Z}$ leads to a smaller F-norm at a high level of obfuscation. This is because $\mathcal{Z}$ corresponds to the size of dataset and is proportional to the number of learning steps per epoch with identical batch sizes, thus reducing the impacts of obfuscation and making $W_{o,5,2}$ and $W_{o,5,3}$ even closer to $W_{o,5,1}$ than $W_{r,5,1}$ is with $T_{o,5,1}$=$T_{r,5,1}$.
\section{Discussions}
\label{sec_discussions}

We discuss challenges of implementing dataset obfuscation, and explore real-world applications particularly on edge networks in this section. We first highlight the following three definitions,
\begin{itemize}
\item \textbf{Privacy}: the degree to which a training dataset $T_{tr}$ is confused with $T_{tr}'$ to prevent data leakage;

\item \textbf{Utility}: the degree to which the obfuscated dataset $T_{tr}'$ can be utilized to allow the local model to achieve acceptable model prediction accuracy upon the testing dataset;

\item \textbf{Distinguishability}: the degree to which the F-norm gap $\Delta$ associated with a raw training dataset $T_{tr}^i$ between its obfuscated dataset $T_{tr}^{i'}$ and a different dataset $T_{tr}^{j'}$ with the same level of obfuscation.
\begin{equation}
\Delta=|\mathcal{D}(W_{T_{tr}^i}, W_{T_{tr}^{i'}})-\mathcal{D}(W_{T_{tr}^i}, W_{T_{tr}^{j'}})|,\ \text{where }i \neq j.
\end{equation}

% \begin{equation}
% \Delta=|\mathcal{D}(W_{T_{tr}^i}, W_{T_{tr}^i'})-\mathcal{D}(W_{T_{tr}^j}, W_{T_{tr}^i'})|,\ \text{where }i \neq j.
% \end{equation}

\end{itemize}

\begin{figure}
	\centering
	\includegraphics[width=0.6\textwidth]{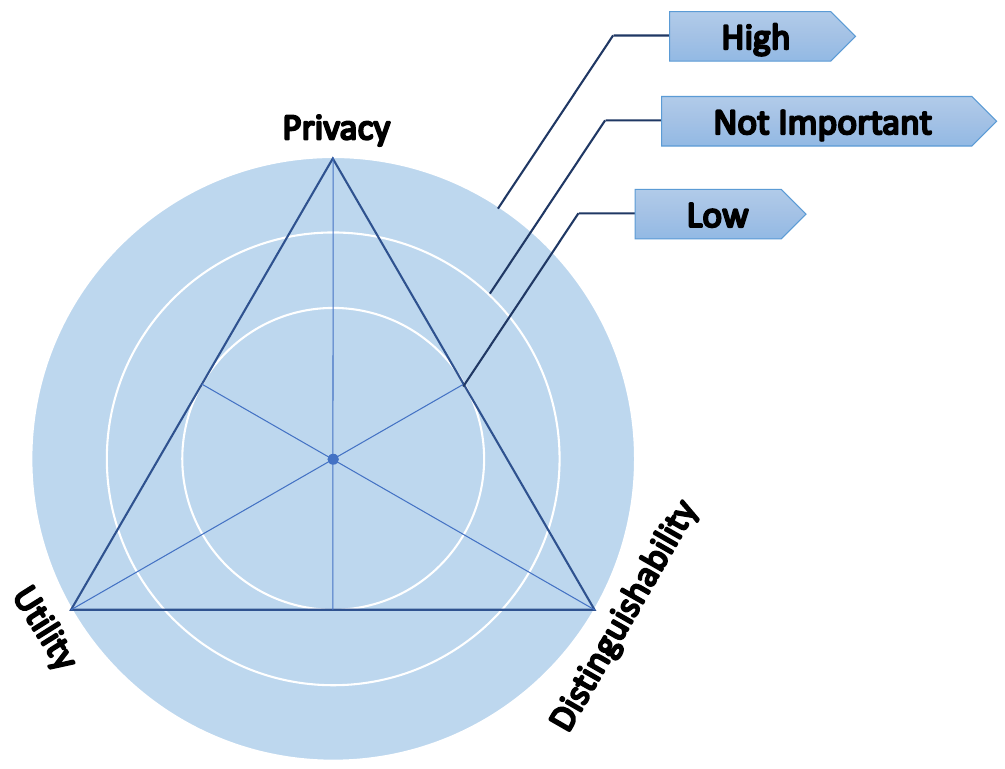}
	\caption{Two examples of the PUD-triangles.}
	\label{fig: pud-triangle}
		\vspace{-1em}  % 调整与下文的间距
\end{figure}

The PUD-triangle refers to the levels in terms of the degree and nature of \textbf{Privacy}, \textbf{Utility}, and \textbf{Distinguishability}, as illustrated in Fig.~\ref{fig: pud-triangle}. We discuss in the following that different applications - examples are given in Sections~\ref{sec_discussion_subsec_outsource} and \ref{sec_discussion_subsec_pol} - satisfy different balance points among the PUD.

% the settlement at a balance point where \textbf{Privacy} and \textbf{Distinguishability} are maximized while \textbf{Utility} can be minimized.

% \begin{figure}[!h]
% 	\centering
% 	\includegraphics[width=0.5\textwidth]{graphs/NoisedML-flowchart.png}
% 	\caption{An illustration of our contribution. We add noise to dataset to preserve privacy and adopt ``F-norm" and ``Model Accuracy" to measure performance. The design goals are making the obfuscation of different datasets distinguishable and preserving the accuracy of the model as much as possible.}
% 	\label{fig: pud-triangle-pol}
% \end{figure}

\subsection{Privacy Decline}
\label{sec_discussion_subsec_privacy}
The more times obfuscated training datasets $T_{tr}'$ are disclosed, the more likely the raw dataset $T_{tr}$ is exposed by simply calculating the average, if the noises are mutually IID among each sharing. One possible solution could be training over multiple fine-grained subsets which are further sampled from the entire local training dataset $T_{tr}$, and applying the non-IID noise along the training process.

\subsection{Training outsourcing on Edge Networks}
\label{sec_discussion_subsec_outsource}

% \begin{figure}[h]
% 	\centering	\includegraphics[width=0.4\textwidth]{graphs/outsourcing.png}
% 	\caption{This figure shows the overview of training outsourcing to cloud platforms.{\color{red}To be removed}}
% 	\label{fig: training-outsouring}
% \end{figure}

Model memorization attacks~\cite{MLMRTM2017} can be exploited in a corrupted training outsourcing task on edge networks. Edge devices with relatively constrained capabilities would have to outsource the training task to the clouds or upper edges by sharing the training dataset. A malicious cloud-based service provider, as the outsourcee, may disclose the raw training dataset $T_{tr}$ to its conspirators by encoding them into the model parameters. Thus, improving the \textbf{Privacy} by the dataset owners (i.e., training outsourcers) obfuscating the dataset before outsourcing them into the machine learning services, as suggested in~\cite{noisy-dataset}. Our experiment results reveal that the model prediction accuracy upon the testing dataset $T_{e}$ is considered acceptable at over 80\% on simple tasks with all samples being obfuscated; see Figs.~\ref{fig:experiment-acc-mnist-degree} and~\ref{fig:experiment-acc-CIFAR-10-degree}.

\subsection{Proof-of-Learning in FL}
% \subsection{PUD-triangle and Proof-of-Learning in Federated Learning}
\label{sec_discussion_subsec_pol}
% We aim to settle at a balance point where \textbf{Privacy} and \textbf{Distinguishability} are maximized while \textbf{Utility} can be minimized. We name it PUD-triangle.

\begin{figure}
	\centering
	\includegraphics[width=0.8\textwidth]{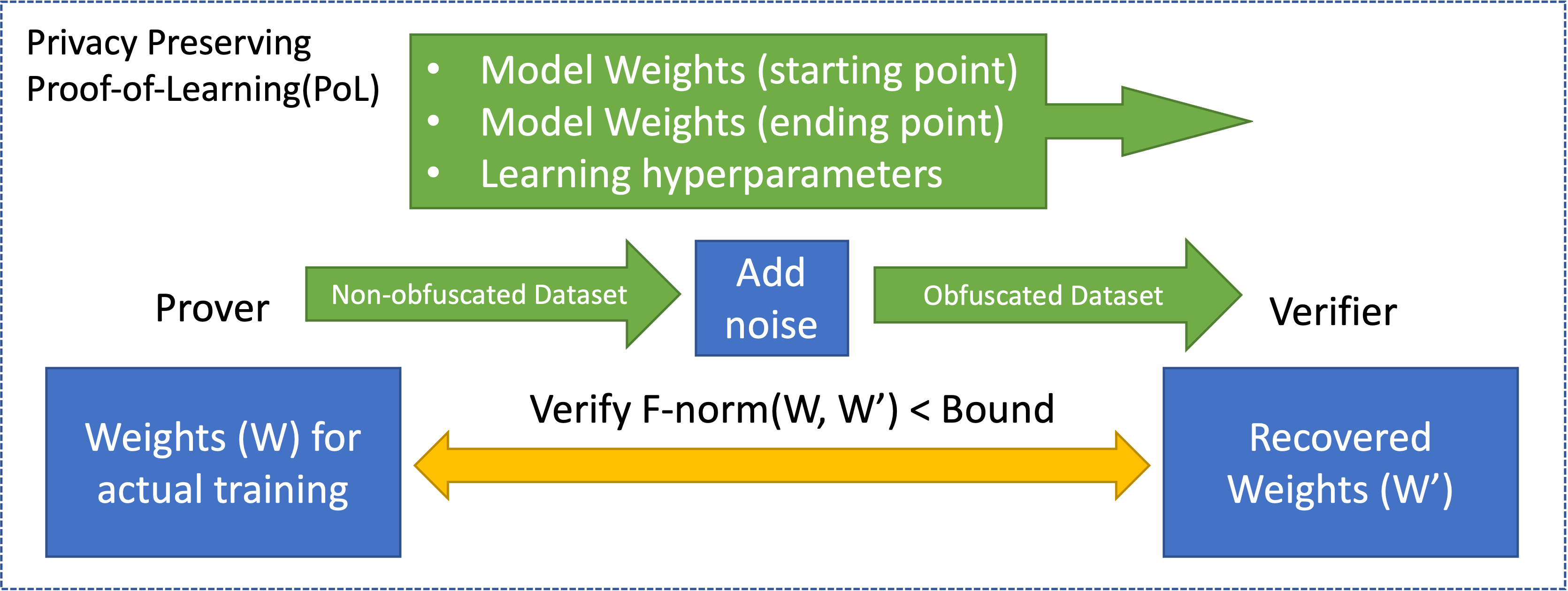}
	\caption{The workflow of Privacy-Preserving PoL.}
	\label{fig: pol}
		\vspace{-1em}  % 调整与下文的间距
\end{figure}

The PUD-triangle is particularly useful in a scenario where the model ownership and the proof of ``works'' for training are both crucial, e.g., the defenses against the model stealing attacks in FL~\cite{9519402}. 
Stealing others' models and faking the ownership can be misleading to those who wish to ensure the necessary training overhead. By launching this attack, a malicious edge device with relatively powerful capability can cheat the whole edge network in regards to the amount of ``works'' that it has done, and steal the rewards. PoL was proposed as a defense strategy against the model stealing attack~\cite{9519402}, which requires the verifiers to replay part of the gradient descent and compare the resultant weights in terms of F-norm $\mathcal{D}$ based on the training dataset $T_{tr}$ shared by the prover (i.e., the owner of the dataset)j; see details in Fig.~\ref{fig: pol}. 

Edge devices implementing PoL in FL tasks have to share the training dataset $T_{tr}$, which is contrary to the original design intent of FL to preserve the privacy of $T_{tr}$. Adding random noises to obfuscate $T_{tr}$ helps preserve privacy. However, this concept comes across two possible issues:
\begin{itemize}
\item Excessive noise added to $T_{tr}^i$ causes a hard differentiation between its obfuscated version $T_{tr}^{i'}$ and an arbitrary obfuscated dataset $T_{tr}^{j'}$. This is known as the strong disguisability of fakes. By sharing a different dataset, a malicious ``prover'' can invalidate the PoL.

\item Too good utility for an obfuscated dataset $T_{tr}'$ can anyhow enrich verifiers' local datasets. Those who abuse the PoL by raising PoL challenges against as many models as possible can easily outperform those who are challenged due to the unbalanced size of the local training dataset under a non-IID setting.
\end{itemize}

We propose to settle at a balanced stable point where spoofing the dataset and abusing the PoL to enrich local resources can be prevented, i.e., maximizing \textbf{Distinguishability} and minimizing the \textbf{Utility}, while featuring a high level of \textbf{Privacy}. Figs.~\ref{fig: experiment-add-noise},~\ref{fig: experiment-accuracy}, \ref{fig:experiment-norm-mnist-overlap} and~\ref{fig:experiment-norm-CIFAR-10-overlap} show the feasibility of meeting the above requirement. 
Adding a certain level of noise to obfuscate the naked recognition can be achieved (Fig.~\ref{fig: experiment-add-noise}, high \textbf{Privacy}), while sharing others the obfuscated dataset during the PoL process should not retain the dataset utility (Fig.~\ref{fig: experiment-accuracy}, low \textbf{Utility}). Otherwise, malicious attackers in FL could just raise PoL challenges at will to enrich their data resources in a non-IID competition. At the same time, the noise level should not be excessive (Figs.~\ref{fig:experiment-norm-mnist-overlap} and~\ref{fig:experiment-norm-CIFAR-10-overlap}, high \textbf{Distinguishability}), or it causes a hard differentiation between an obfuscated dataset and an arbitrary dataset. This indicates that a PoL prover could send PoL verifiers a different dataset and pretend to have the correct dataset. As a result, PoL is invalidated by verifiers having to compare the F-norm between the claimed model and the replayed model based on a different dataset. This has been adopted in emerging FL frameworks~\cite{yu2023ironforge}.

% see Section~\ref{subsec: experiement results}.

% %=================================================   
% %=================================================   
% \section{Introduction}
% %=================================================   

% %=================================================   
% \section{Related Work}
% \label{sec-relatedwk}
% %===============================================

% %=================================================  
\section{Conclusion}
\label{sec-conclu}
We conducted comprehensive experiments to investigate
% We conducted a more comprehensive experiment than existing studies. The experiment investigated 
how obfuscating a training dataset by adding random noises would affect the resultant model in terms of model prediction accuracy, model distance, data utility, data privacy, etc. Experimental results over both the MNIST and CIFAR-10 datasets revealed that, if a balanced point among all the considered metrics is satisfied, broad application prospects on real-world applications will become realistic, including training outsourcing on edge computing and guarding against attacks in FL among edge devices.
% %=================================================   

% \medskip
% \noindent\textbf{Acknowledgement.}

%================================================
%=================================================
% \newpage
\bibliographystyle{unsrt}
\bibliography{bib.bib}

\end{document}